\documentstyle[12pt]{article}
\newcommand{\Tr}[1]{\,{\rm Tr}\,#1\,}
\begin{document}
\title{
\begin{flushright}
{\small SMI-3-95}
\end{flushright}
\vspace{0.5cm}
Bosonization of fermion determinants} \author{ A.A.Slavnov \thanks{E-mail:$~~$
slavnov@class.mian.su} \\ Steklov Mathematical Institute, Russian
Academy of Sciences,\\ Vavilov st.42, GSP-1,117966, Moscow, Russia }
\maketitle

\begin{abstract}

A four dimensional fermion determinant is presented as a path integral of
the exponent of a local five dimensional action describing
constrained bosonic system. The construction is carried out
both in the continuum theory and in the lattice model. \end{abstract}

\section {Introduction}

The problem of bosonization of fermionic theories was studied by many
authors (see e.g. \cite{C}, \cite{PW}, \cite{W}, \cite{GSS}, \cite{FGS},
\cite{N}), mainly in the context of two-dimensional models. There were
also some attempts to extend this procedure to higher dimensions
\cite{Sem}, \cite{ML1}, \cite{M}, \cite{Z}, but at present they lead only
to partial success.

The problem of bosonization is of a particular importance for lattice
gauge theori\-es as lattice QCD, because the fermionic degrees of freedom
complicate numeri\-cal simulations enormously.

Recently M.Lusher \cite{ML} proposed the algorithm for the approximate
inversion of the QCD fermion determinant replacing it by an infinite
series of bosonic determi\-nants. In the present paper I will describe an
alternative approach which allows to write the exact expression for the
fermion determinant as a path integral of the exponent of a local bosonic
action. In my approach a four dimensional fermionic system is
replaced by a five dimensional constrained bosonic one.

In the second section I present the construction for the continuum
theory. In this case the proof of equivalence will be given in the
framework of perturbation theory as the problem of nonperturbative
definition of a continuum path integral is at present beyond our
possibilities. In the third section the corresponding procedure will
be given for the lattice QCD where the equivalence may be proven at
the nonperturbative level as well.
\section {Perturbative bosonization in the continuum theory}

We consider the following fermion determinant
\begin{equation}
\det(D+m)^n= \int \exp \{- \sum_{i=1}^n \int d^4x[ \bar{
\psi}^i_{\alpha}(x)(D_ {\alpha \beta}+m \delta_{\alpha \beta})
\psi^i_{\beta}(x)] \}d \bar{\psi}^i_{\alpha}d \psi^i_{\beta} \label{1}
\end{equation} where \begin{equation} D_{\alpha \beta}=( \gamma_{\mu}
\partial_{\mu}+ig \gamma_{\mu}A_{\mu})_{\alpha \beta} \label{2}
\end{equation} $A_{\mu}$ belongs to the Lie algebra of some compact group
($SU(3)$ for QCD).  Indices $ \alpha$ denote spinorial and colour
 components. The eq.( \ref{1}) describes $n$ degenerate fermion flavours
interacting vectorialy with the Yang-Mills field. To make contact with the
discussion of lattice models we assume that the four dimensional space is
Euclidean and the $ \gamma_{\mu}$ matrices ($ \mu=1,2 \ldots 5$) are
Hermitean.  As all the discussion in this section will be restricted to
perturbation theory all the equations are understood as a series over $g$,
and the Gaussian path integrals are defined axiomatically as in ref.
\cite{S}, \cite{SF}.  Therefore we need not to bother about the
convergence of the path integrals and all the conclusions are valid
(perturbatively) for any number of fermion flavours.  However in the next
section where we deal with the nonperturbative proof for lattice models we
have to be more careful at this point and to make all the integrals
convergent we restrict our discussion to the case of even number of
flavours. For simplicity we take $n=2$.

We present the determinant of the Dirac operator as the determinant of
the Hermitean operator by using the identity \begin{equation} \det(
D+m)= \det[\gamma_5(D+m)]
\label{3} \end{equation} Therefore \begin{equation} \det(D+m)^2= \det(
\gamma_5(D+m) \gamma_5( D+m))= \det(-D^2+m^2) \label{4} \end{equation}
Being the square of the Hermitean operator the operator $-D^2+m^2$ is
positive definite.

Let us introduce the following action depending on five
dimensional bosonic fields $ \phi_{\alpha}(t,x)$ \begin{equation} S= \int
dtd^4x\{ \lambda \phi^*_{\alpha}(t,x) \partial_t \phi_{\alpha}(t,x)+
\phi^*_{\alpha}(t,x)[-D^2(x)+m^2]_{\alpha \beta} \phi_{\beta}(t,x) \}
\label{5} \end{equation} Here $ \lambda$ is a small positive parameter
which eventually will be put equal to zero. The fields $ \phi_{\alpha}$
have the same spinorial and colour structure as the fields $
\psi_{\alpha}$.

We claim that the determinant (\ref{1}) (for $n=2$) may be presented as
the following path integral:  \begin{eqnarray} \det(D+m)^2= \lim_{\lambda
\rightarrow 0} \int \exp \{-S+i \int_0^1 dt \int d^4x \phi^*_{\alpha}(x,t)
\chi_{\alpha}(x) + \nonumber \\+i \int_0^1dt \int d^4x \chi^*_{\alpha}(x)
\phi_{\alpha}(x,t) \}d \phi^*_{\alpha}d \phi_{\beta}d \chi_{\alpha}^*d
\chi_{\beta} \label{6} \end{eqnarray} The fields $ \chi(x)$ play a role
of Lagrange multipliers imposing the constraints \begin{equation} \int_0^1
dt \phi^*(t,x)= \int_0^1 dt \phi(t,x)=0 \label{7} \end{equation}  The
Green functions of the fields $ \phi$ defined by the action(\ref{5}) are
retarded:  \begin{equation} \tilde{G}(k,p)=
\frac{1}{i \lambda k+p^2+m^2} \label{8} \end{equation}
\begin{equation} G(t,p)= \frac{1}{2 \pi} \int \exp \{ikt \}
\tilde{G}(k,p)dk;\quad G(t)=0, t<0 \label{9} \end{equation} Integrating
in the eq.(\ref{6}) over $ \phi^*, \phi$ one gets \begin{equation} I=
\det( \lambda \partial_t -D^2+m^2)^{-1} \times \label{10} \end{equation}
 $$ \int \exp \{- \int_0^1dt ds \int d^4x d^4y \chi_{\alpha}^*(x)[ \lambda
 \partial_t -D^2+m^2]^{-1}_{ \alpha \beta}(x,y,s-t)
\chi_{\beta}(y) \}d \chi^* d \chi $$ Due to the fact that the Green
functions of the fields $ \phi$ are retarded, for any $ \lambda$ different
from zero the first factor in the eq.( \ref{10}) is equal to one.  Indeed
a perturbative expansion of this determinant generates the diagrams of the
type \begin{equation} \Pi_n \sim \int dk \frac{F(k,p_1 \ldots p_n)}{(i
\lambda k+p_1^2+m^2) \ldots (i \lambda k+p_n^2+m^2)} \label{11}
\end{equation} where $F$ is some polynomial over $p_i,k$. Note that the
fields $A$ do not depend on $s$, therefore the diagrams generated by
eq.(\ref{10}) are the cycles with zero fifth component of momenta. The
integral over $k$ is convergent and all the poles are in the upper complex
half plane.  Closing the integration contour in the lower half plane one
sees that the integral is equal to zero. In fact one has to be slightly
more careful at this point, as due to momentum conservation all these
diagrams are multiplied by the singular constant factor $ \sim \delta
(0)$. To give a precise meaning to this expression some infrared cut-off
must be introduced to the eq.  ( \ref{5}). In the next section we shall
show that a finite lattice may serve this purpose.

The second factor is not singular in the limit $
\lambda \rightarrow 0$ and one can calculate it by putting $ \lambda =0$.
Let us prove it in some more details.

Expansion of the exponent in the eq.(\ref{10}) over $g$ generates the
terms of the form \begin{equation} \int_0^1 ds \int_0^1 dt \int dk \frac{
\exp \{ik(s-t) \}P( \lambda k,p_1 \ldots p_n)}{(i \lambda k+p_1^2+m^2)
\ldots (i \lambda k+p_n^2+m^2)} \label{12} \end{equation}
where $P$ is some polynomial over $k,p_i$.
 Integrating over $s$ and $t$ one gets
\begin{equation}
\int dk \frac{(2-e^{-ik}-e^{ik})P( \lambda k,p_i)}{(k-i \epsilon)^2(i
\lambda k+p_1^2+m^2) \ldots (i \lambda k+p_n^2+m^2)} \label{13}
\end{equation}
The pole at $k=0$ is spurious. It is compensated by the zero of the
nominator. So we are free to define the roundabout of this pole
arbitrarily. We choose the prescription $k \rightarrow k-i \epsilon$.
With this prescription the part of the integrand proportional to $(2-
\exp \{-ik \})$ gives zero contribution as in this case we may close the
integration contour in the lower half plane and all the poles are situated
in the upper half plane. The last term, proportional to $ \exp \{ik \}$ is
calculated by closing the contour in the upper half plane. The integral is
equal to the sum of the residues at the poles at $k=0, k= i
{\lambda}^{-1} (p_i^2+m^2)$. The contribution of the poles
at $k=i{\lambda}^{-1} (p_i^2+m^2)$ vanishes at $ \lambda \rightarrow 0$.
 The contribution of the pole at $k=0$ gives
\begin{equation} \frac{P(k,p_i)}{(i \lambda k+p_1^2+m^2) \ldots (i
\lambda k+p_n^2+m^2)}|_{k=0} \label{14}
\end{equation}
Therefore the integral (\ref{10}) acquires a form
\begin{equation}
I=\int \exp \{- \int  \chi_{\alpha}^*(x)[-D^2+m^2]^{-1}_{\alpha \beta}(x,y)
\chi_{\beta}(y)dxdy+O(
\lambda)\}d \chi_{\alpha}^*d \chi_{\beta} \label{15}
\end{equation}
Integrating over $ \chi^*, \chi$ we get
\begin{equation}
I= \det(-D^2+m^2)+O( \lambda) \label{16}
\end{equation}
Therefore
\begin{equation}
\lim_{ \lambda \rightarrow 0}I= \det(D+m)^2 \label{17}
\end{equation}

\section {Bosonization on the lattice}

In this section we generalize the construction given above to the
case of lattice models, where the proof of equivalence can be given
without any references to perturbation theory.

We again present the determinant of the Dirac operator as the
determinant of the Hermitean operator by using the identity
\begin{equation}
\det( \hat{D}+m)= \det[\gamma_5( \hat{D}+m)], \quad \hat{D}=
\gamma_{\mu}D_{\mu} \label{18} \end{equation}
In eq.( \ref{18}) $D_{\mu}$ is the lattice covariant derivative
\begin{equation}
D_{\mu} \psi(x)= \frac{1}{2a}[U_{\mu}^+(x) \psi(x+a_{\mu})-U_{\mu}(x)
\psi(x-a_{\mu}] \label{19}
\end{equation}
$U_{\mu}$ is a lattice gauge field. We consider a finite lattice with
periodic boundary conditions.

As it has been already mentioned, to provide the positivity of
the determinant we consider the case of two degenerate fermion flavours
interacting vectorialy with the Yang-Mills field. We shall not deal
explicitely with the problem of fermion doubling.  All the reasonings are
trivially extended to the model improved by adding for example the Wilson
term.

It is convinient to present the fermion determinant in the following form
\begin{equation}
\det[\gamma_5( \hat{D}+m)]^2= \int \exp \{a^4 \sum_x
\bar{\psi}(x)( \hat{D}^2-m^2) \psi(x) \} d \bar{ \psi}d \psi \label{20}
\end{equation}

We shall prove that in analogy with the continuum case the integral
over fermionic fields $ \psi$ can be replaced by an integral
of a five dimensional lattice bosonic action. The spatial
components $x$ are defined as above. The fifth component $t$ to be defined
on the one dimensional lattice of the length $L$ with the lattice spacing $b$:
\begin{equation}
L=2Nb, \quad -N<n \leq N \label{21}
\end{equation}
We choose $b$ in such a way that $b<<a$ and in the continuum
limit
\begin{equation}
2Nb^2=Lb \rightarrow 0 \label{22}
\end{equation}
The key ingredient of the construction presented in the preceeding section
was the using of retarded Green functions for the fields $
\phi$ providing triviality of the determinant in the eq.(\ref{10}). This
phenomenon is a continuum analog of the well known property of a
triangular matrix: the determinant of a triangular matrix is equal to the
product of the diagonal elements. Therefore to achieve the same goal in
the lattice case we choose the lattice derivative with respect to the
fifth coordinate $t$ in the form of a triangular matrix. More precisely we
are going to prove the following equality \begin{equation} \int \exp \{a^4
\sum_x \bar{ \psi}(x)( \hat{D}^2-m^2) \psi(x) \}d \bar{\psi}d \psi =
\label{23} \end{equation} $$ = \lim_{\lambda \rightarrow 0,b \rightarrow
0} \int \exp \{a^4b \sum_{n=-N+1}^N \sum_x [ \lambda \frac{
\phi^*_{n+1}(x)- \phi^*_n(x)}{b} \phi_n(x)+ \phi^*_n(x)( \hat{D}^2-m^2)
\phi_n(x)- $$ $$ - \frac{i}{ \sqrt{L}}( \phi^*_n(x) \chi(x)+ \chi^*(x)
\phi_n(x))] \}d \phi^*_nd \phi_nd \chi^*d \chi .  $$ Here $ \phi_n(x) $
and $ \chi(x)$ are bosonic fields which carry the same spinorial and
colour indices as the fields $ \psi(x)$. The fields $
\phi_n(x)$ satisfy free boundary conditions i.e.  \begin{equation}
\phi_n=0, \quad n \leq -N, \quad n>N \label{24} \end{equation}

The operator $- \hat{D}^2+m^2$ is Hermitean and can be diagonalized by a
unitary transformation. As it does not depend on $t$ this transformation
will make the exponent in the r.h.s. of eq.( \ref{23}) diagonal with
respect to all variables except for $t$. After this transformation
the r.h.s. of eq.( \ref{23}) acquires the form
$$
I= \lim_{ \lambda \rightarrow 0, b \rightarrow 0}I( \lambda, b)
$$
\begin{eqnarray}
I( \lambda,b)= \int \exp \{b \sum_{n=-N+1}^N \sum_{ \alpha}[ \lambda
\frac{ \phi^{* \alpha}_{n+1}- \phi^{* \alpha}_n}{b} \phi^{ \alpha}_n -
\phi^{* \alpha}_nB^{\alpha} \phi^{\alpha}_n \nonumber \\ - \frac{i}{
\sqrt{L}}( \phi^{* \alpha}_n \chi^{\alpha}+ \chi^{* \alpha}_n \phi^{
\alpha}_n)] \}d \phi^*_n d \phi_n d \chi^*d \chi; \quad \phi^{*
\alpha}_{N+1}=0 \label{25} \end{eqnarray} Index $ \alpha$ refers now to
the eigenstates of the operator $- \hat{D}^2+m^2$, $B_{\alpha}$ being the
corresponding eigenvalues.  Obviously $B_{\alpha}>0$.

The integral ( \ref{25}) is convergent as the real part of the action in
the exponent is positive. It is easy to see by rewriting the action in
terms of Fourier components:
\begin{eqnarray}
S= \frac{1}{2 \pi} \int_{- \frac{ \pi}{b}}^{ \frac{ \pi}{b}} \{ \tilde{
\phi}^{* \alpha}(k)[- \lambda(e^{-ikb}-1)b^{-1}+B^{ \alpha}] \tilde{
\phi}^{\alpha}(k)]+ \nonumber\\
+ \frac{i}{2 \pi \sqrt{L}}(e^{-iknb} \tilde{ \phi}^{* \alpha}(k) \chi^{
\alpha}+e^{iknb} \chi^{* \alpha} \tilde{ \phi}^{\alpha}(k))
\}dk  \label{26} \end{eqnarray}
The real part is
\begin{equation}
Re S= \frac{1}{2 \pi} \int_{- \frac{ \pi}{b}}^{ \frac{ \pi}{b}} \{
\tilde{ \phi}^{* \alpha}(k)[- \lambda(
\cos{kb}-1)b^{-1}+B^{\alpha}] \tilde{\phi}^{\alpha}(k)]dk \label{27}
\end{equation}
As $ \lambda<0$ and $B^{\alpha}>0$, $Re S>0$.

Performing in the eq.( \ref{25}) the integration over $ \phi_n$ one gets
\begin{equation} I( \lambda,b)= \det(C^{-1}) \int \exp
\{ \frac{-b}{L} \sum_{\alpha} \sum_{n,m=-N+1}^N \chi^{*
\alpha}(C^{\alpha})^{-1}_{mn} \chi^{\alpha} \} d \chi^* d \chi \label{28}
\end{equation}
Where $C_{\alpha}(k)$ is the kernel of the quadratic form in the eq.(
\ref{25}).

As follows from the eq.( \ref{25})
\begin{equation}
\det(C)= \prod_{\alpha} \det(C_{\alpha}) \label{29}
\end{equation}
In the coordinate space $C_{\alpha}$ is a triangular
matrix with the diagonal elements \begin{equation} \sim ( \lambda
+B_{\alpha}b) \label{30} \end{equation} Therefore \begin{equation}
\det(C)= \exp \{ \sum_{\alpha} \ln( \lambda+B_{\alpha}b)^{2N} \}
 \label{31} \end{equation} Separating the constant term one gets
\begin{equation} \det(C)= \exp \{2Nb \lambda^{-1} \sum_{\alpha}B_{\alpha}
+O(Lb) \}= \label{32} \end{equation} $$ = \exp \{ \lambda^{-1}L \Tr[-
\hat{D}^2+m^2]+O(Lb) \} $$ Using the explicit form of the operator $D$ one
can easily verify that $ \Tr[- \hat{D}^2+m^2]$ is a nonessential constant.
It follows also from the fact that the trace of a local operator is local.
 It has to be gauge invariant and a polynomial of the second order
in the fields $A_{\mu}$. The only possible solution is a constant. So we
can include $ \det(C^{-1})$ into normalization constant.

To get an explicit form of the remaining terms it is sufficient to find
the stationary point of the exponent in the eq.( \ref{25}). The
corresponding classical equations look as follows
\begin{equation}
- \lambda( \phi^{* \alpha}_{n+1}- \phi^{*\alpha}_n)b^{-1}+B^{\alpha}
\phi^{* \alpha}_n +iL^{- \frac{1}{2}} \chi^{* \alpha}=0; \quad n \neq N
\label{33}
\end{equation}
$$
 \lambda( \phi^{\alpha}_n- \phi^{\alpha}_{n-1})b^{-1}+B^{\alpha}
 \phi^{\alpha}_n+iL^{- \frac{1}{2}} \chi^{\alpha} =0; \quad n \neq-N+1
$$
\begin{equation}
- \phi^{* \alpha}_Nb^{-1} \lambda- \phi^{* \alpha}_NB^{\alpha}-iL^{-
\frac{1}{2}} \chi^{* \alpha}=0 \label{34}
\end{equation}
$$
 \phi^{ \alpha}_{-N+1}b^{-1} \lambda+ \phi^{
\alpha}_{-N+1}B^{\alpha}+iL^{- \frac{1}{2}} \chi^{ \alpha}=0
$$
for small $b$ eq.s( \ref{33}) may be approximated by the differential
equations
\begin{equation}
- \lambda \partial_t \phi^{* \alpha}+B^{\alpha} \phi^{* \alpha}+iL^{-
\frac{1}{2}} \chi^{* \alpha}=0 \label{35} \end{equation}  $$  \lambda
\partial_t \phi^{ \alpha}+B^{\alpha} \phi^{\alpha}+iL^{- \frac{1}{2}}
\chi^{ \alpha}=0 $$
whereas eq.s( \ref{34}) play the role of boundary conditions:
\begin{equation}
\phi^{* \alpha}( \frac{L}{2})=0; \quad \phi^{\alpha}(- \frac{L}{2})=0
\label{36} \end{equation} The solution of these eq.s is \begin{equation}
\phi^{* \alpha}(t)=- \frac{i}{ \sqrt{L}B^{\alpha}} \chi^{* \alpha}(1- \exp
\{B^{\alpha} \lambda^{-1}(t- \frac{L}{2}) \}) \label{37}
\end{equation}
$$
\phi^{ \alpha}(t)=- \frac{i}{ \sqrt{L}B^{\alpha}} \chi^{ \alpha}(1- \exp
\{-B^{\alpha} \lambda^{-1}(t+ \frac{L}{2}) \})
$$
Substituting these solutions to the eq.( \ref{25}) we get in the limit $b
\rightarrow 0$
\begin{equation}
\lim_{b \rightarrow 0}I( \lambda,b)= \int \exp \{- \int_{- \frac{L}{2}}^{
\frac{L}{2}} \sum_{\alpha} \frac{\chi^{* \alpha} \chi^{
\alpha}}{B^{\alpha}L}[1- \exp \{ \frac{B^{\alpha}}{ \lambda}(t-
\frac{L}{2}) \}]dt \}d \chi^*d \chi = \label{38} \end{equation} $$ = \int
\exp \{- \sum_{\alpha} \chi^{* \alpha}(B^{\alpha})^{-1} \chi^{\alpha}[1-
\frac{ \lambda}{B^{ \alpha}L}[1+ \exp \{-B^{\alpha} \lambda^{-1}L \}]] \}d
\chi^*d \chi $$ In the limit $ \lambda \rightarrow 0$ this equation
reduces to \begin{equation} I= \lim_{\lambda \rightarrow 0,b \rightarrow
0}I( \lambda,b)= \exp \{- \sum_{\alpha} \chi^{* \alpha}(B^{\alpha})^{-1}
\chi^{\alpha} \}d \chi^{* \alpha} d \chi^{\alpha} = \det(- \hat{D}^2+m^2)
\label{39} \end{equation} The equality (\ref{23}) is proven. The final
recipe is the following:  Any gauge invariant observable may be expressed
as a purely bosonic path integral of the form \begin{eqnarray} Z= \lim_{
\lambda \rightarrow 0, b \rightarrow 0} \int \exp \{a^4b \sum_{n=-N+1}^N
\sum_x [ \lambda \frac{ \phi^*_{n+1}(x)- \phi^*_n(x)}{b} \phi_n(x)+
\nonumber \\+ \phi^*_n(x)( \hat{D}^2-m^2) \phi_n(x)- \frac{i}{ \sqrt{L}}(
\phi^*_n(x) \chi(x)+ \chi^*(x) \phi_n(x))]+ \nonumber \\+L(U)+s.t.  \}d
\phi^*_nd \phi_n d \chi^*d \chi \label{36} \end{eqnarray} where $L(U)$ is
the lattice Yang-Mills action and s.t. stands for the source term
depending on the fields $U_{\mu}, \phi^*_n, \phi_n$. All the fields are
bosonic and the integration goes over the fields $ \phi^*_n, \phi_n$
defined on the one-dimensional lattice with length $L$ and free boundary
conditions.

\section {Discussion}

In this paper we showed that a four dimensional fermion determinant can be
written as a path integral of the exponent of a five dimensional local
bosonic action. In the same way one can present a two dimensional fermion
determinant as a path integral for three-dimensional bosonic theory. In
the case of lattice models this procedure leads to a well defined
bosonic path integral. No numerical simulations in this approach have
been tried so far and it would be very important to see how the method
works in practical calculations. In this case one
need not of course to take the limit $ \lambda \rightarrow 0, b
\rightarrow 0$, but the following conditions $b<<a, \quad b<< \lambda,
\quad b<<L^{-1}$ must be respected.

{\bf Acknowledgements.} \\
This work was started while the author was visiting
Universite Libre de Bruxelles and The Institute for Theoretical Physics of
University of California at Santa Barbara, participating in the Workshop
on Chiral symmetry on the lattice organised by M.Creutz. I am grateful to
Marc Henneaux for hospitality at ULB and useful discussions. I also
wish to thank the staff of ITP for hospitality and all the
participants of the chiral workshop for numerous stimulating
discussions. I am indebted to S.A.Frolov for helpful comments.
This researsh was supported in part by International Science Foundation
and Russian Governement under grant MNB000, Russian Basic Research Fund
under grant 94-01-00300a and the National Science Foundation under Grant
PHY89-04035.$$ ~ $$  
\begin{thebibliography}{99} {\small \bibitem{C}
S.Coleman, Phys.Rev.D11 (1975) 2088.  \bibitem{PW} A.Polyakov and
P.Wigman, Phys.Lett.B131 (1983) 121; 141 (1984) 223.  \bibitem{W}
E.Witten, Comm.Math.Phys.92 (1984) 455.  \bibitem{GSS} R.Gamboa-Saravi,
F.Shaposnik and J.Solomin, Nucl.Phys.B185 (1981) 239.  \bibitem{FGS}
K.Furuya, R.Gamboa-Saravi and F.Shaposnik, Nucl.Phys.B208 (1982) 159.
\bibitem{N} C.Naon, Phys.Rev. D31 (1985) 2035.  \bibitem{Sem}
G.W.Semenoff, Phys.Rev.Lett.61 (1988) 817.  \bibitem{ML1} M.Lusher,
Nucl.Phys. B326 (1989) 557. \bibitem{M} E.C.Marino, Phys.Lett. B263 (1991)
63. \bibitem{Z} L.Huerta,F.Zanelli, Phys.Rev.Lett.  71 (1993) 3622.
\bibitem{ML} M.Lusher Nucl.Phys.  B418 (1994) 637.  \bibitem{S}
A.A.Slavnov, Theor.Math.Phys. 22 (1975) 177.  \bibitem{SF} L.D.Faddeev,
A.A.Slavnov,{\it Gauge Fields.  Introduction to Quantum Theory.} Second
edition.  Moscow, Nauka, 1988.  English translation ed.Addison-Wesley 1990
} \end {thebibliography}
\end{document}